\documentclass[reprint,aps,prb,longbibliography,showpacs,amsmath,amssymb,superscriptaddress]{revtex4-2}
\usepackage{booktabs}
\usepackage{hhline}
\usepackage{amsmath}
\usepackage{amssymb}
\usepackage[version=4]{mhchem}
\usepackage{siunitx}
\DeclareSIUnit\angstrom{\text{\AA}}

\usepackage{svg}
\usepackage{graphicx}
\usepackage{dcolumn}
\usepackage{bm}

\usepackage{hyperref}
\hypersetup{
    unicode=false,
    pdftoolbar=true,
    pdfmenubar=true,
    pdffitwindow=false,
    pdfstartview={FitH},
    colorlinks=true,
    linkcolor=blue,
    citecolor=blue,
    urlcolor=blue,
}

\usepackage{mathptmx}
\usepackage{etoolbox}

\graphicspath{ {./images/}}
\usepackage{float}
\usepackage{multirow}

\pdfoutput=1

\begin{document}

\title{High-performance training and inference for deep equivariant interatomic potentials}

\newcommand\HARVARD{John A. Paulson School of Engineering and Applied Sciences,
Harvard University, Cambridge, MA, USA}

\author{Chuin Wei Tan}
\thanks{Co-first author}
\affiliation{\HARVARD}

\author{Marc L. Descoteaux}
\thanks{Co-first author}
\affiliation{\HARVARD}

\author{Mit Kotak}
\affiliation{Center for Computational Science and Engineering, Massachusetts Institute of Technology, Cambridge, MA, USA}

\author{Gabriel de Miranda Nascimento}
\affiliation{Department of Materials Science and Engineering, Massachusetts Institute of Technology, Cambridge, MA, USA}

\author{Seán R. Kavanagh}
\affiliation{Center for the Environment, Harvard University, Cambridge, MA, USA}

\author{Laura Zichi}
\affiliation{\HARVARD}

\author{Menghang Wang}
\affiliation{\HARVARD}

\author{Aadit Saluja}
\affiliation{\HARVARD}

\author{Yizhong R. Hu}
\affiliation{\HARVARD}

\author{Tess Smidt}
\affiliation{Research Laboratory of Electronics, Massachusetts Institute of Technology, Cambridge, MA, USA}

\author{Anders Johansson}
\affiliation{Sandia National Laboratories, Albuquerque, NM, USA}

\author{William C. Witt}
\affiliation{\HARVARD}

\author{Boris Kozinsky}
\affiliation{\HARVARD}
\affiliation{Robert Bosch LLC Research and Technology Center, Watertown, MA, USA}

\author{Albert Musaelian}
\thanks{Contact: \href{mailto:allegro-nequip@g.harvard.edu}{allegro-nequip@g.harvard.edu}}
\affiliation{\HARVARD}
\affiliation{Mirian Technologies Inc., Boston, MA, USA}

\begin{abstract}
    Machine learning interatomic potentials, particularly those based on deep equivariant neural networks, have demonstrated state-of-the-art accuracy and computational efficiency in atomistic modeling tasks like molecular dynamics and high-throughput screening. 
    The size of datasets and demands of downstream workflows are growing rapidly, making robust and scalable software essential.
    This work presents a major overhaul of the NequIP framework focusing on multi-node parallelism, computational performance, and extensibility.
    The redesigned framework supports distributed training on large datasets and removes barriers preventing full utilization of the PyTorch 2.0 compiler at train time.
    We demonstrate this acceleration in a case study by training Allegro models on the SPICE 2 dataset of organic molecular systems. 
    For inference, we introduce the first end-to-end infrastructure that uses the PyTorch Ahead-of-Time Inductor compiler for machine learning interatomic potentials.
    Additionally, we implement a custom kernel for the Allegro model's most expensive operation, the tensor product.
    Together, these advancements speed up molecular dynamics calculations on system sizes of practical relevance by up to a factor of 18.
\end{abstract}

\maketitle

\section{Introduction}

Machine learning interatomic potentials (MLIPs) have become an essential tool for computational materials science and chemistry \citep{PhysRevLett.98.146401, PhysRevLett.104.136403}. 
Their widespread adoption has driven the development of specialized software frameworks for their training and inference \citep{schutt2023schnetpack, zeng2023deepmd, lysogorskiy2021performant, fan2022gpumd, witt2023acepotentials, salzbrenner2023developments, podryabinkin2023mlip, pelaez2024torchmd, firoz2025optimizing}.
Meanwhile, the growing availability of large and diverse datasets, such as SPICE \citep{eastman2023spice}, MPTrj \citep{deng2023chgnet}, Alexandria \citep{schmidt2024improving}, OMat24 \citep{barroso2024open}, and MatPES \citep{kaplan2025foundational}, has enabled the development of more generalizable MLIPs capable of serving as pretrained potentials for a wide range of applications \citep{chen2022universal, deng2023chgnet, kovacs2023mace, batatia2023foundation, merchant2023scaling, park2024scalable, eastman2024nutmeg, yang2024mattersim, neumann2024orb, rhodes2025orb, fu2025learning}, including the prediction of ionic conductivity in battery cathodes \citep{deng2023chgnet}, thermal properties via molecular dynamics and phonon calculations \cite{kaplan2025foundational}, and defect geometries in inorganic solids \citep{kavanagh2024identifying, mosquera2024machine, berger2025screening}.
In this rapidly evolving landscape, scalable, efficient, and adaptable software infrastructure is essential. Frameworks must not only handle the computational demands of large datasets but also support novel training paradigms, facilitate efficient hyperparameter optimization, and integrate with diverse hardware environments.  \\

This work presents a major revamp of the NequIP software framework for MLIPs to dramatically improve computational performance and versatility across training and inference tasks of all sizes.
This framework was originally developed to support the NequIP \citep{batzner20223} and Allegro \citep{musaelian2023learning} model architectures, which are deep equivariant neural networks that incorporate rotational, mirror, and other physical symmetries directly in the structure of the model.
Equivariant model architectures have been shown to improve data efficiency, enhance generalization, and achieve state-of-the-art accuracy in MLIP-based simulations \citep{batzner20223, musaelian2023learning, batatia2022mace, PhysRevX.14.021036}.
Equivariant neural networks, like many MLIPs, pose practical computational challenges due to their use of mathematical operations for which libraries like PyTorch do not provide optimized implementations.
More powerful compilers and custom optimized implementations can overcome these challenges, motivating this work.  \\

We begin in Section~\ref{methods} by describing the methods we used to accelerate training and inference. Then, we present a case study demonstrating the performance improvements using the Allegro deep equivariant network architecture \citep{musaelian2023learning} in Section~\ref{results}.

\section{Training and Inference Acceleration}
\label{methods}

\subsection{TorchInductor Compilation}
\label{inductor}

The NequIP framework is built on top of the PyTorch neural network library \citep{paszke2019pytorch}.
PyTorch 2.0 introduced TorchInductor, a compiler that transforms graphs of high-level PyTorch operations into highly-optimized, device-specific code \citep{ansel2024pytorch}.
On GPUs, TorchInductor leverages Triton \citep{tillet2019triton}, a hardware-agnostic language, to generate optimized kernels, while on CPUs, it produces C++ and OpenMP code \citep{dagum1998openmp}. 
TorchInductor can apply significantly more complex kernel fusions than the TorchScript just-in-time compiler \citep{ansel2024pytorch} previously used by the NequIP framework.
Its advanced fusion, inlining, and loop/layout reordering features have led to substantial performance improvements in many applications \citep{hao2023torchbench, zhu2024scalefold, kasimbeg2025accelerating}.  \\

The main challenge in applying TorchInductor to NequIP and Allegro is the models' use of automatic differentiation to predict forces and virials as exact derivatives of the potential energy. 
Enforcing these derivative relationships ensures that forces remain conservative fields of the energy, which is essential for maintaining energy conservation in molecular dynamics simulations \citep{fu2022forces}.
Evidence suggests that using conservative forces improves the accuracy of physical property predictions \citep{fu2025learning}, particularly for phonon-related properties \citep{loew2024universal}, and enhances the stability of geometry relaxation tasks \citep{bigi2024dark}.
Computing the derivative, however, is not a primitive PyTorch operation that is supported directly by the compiler machinery.
To overcome this challenge, we adapted our code to trace the entire model, including the calculation of the derivative(s), into a single computational graph that is represented using PyTorch's \texttt{torch.fx} library \citep{reed2022torch}.
The \texttt{torch.fx} graph resulting from this procedure represents both the forward pass, which computes the potential energy, and the backward-mode automatic differentiation pass, which computes forces and other derivatives. Because this graph contains only primitive PyTorch operations, it is compatible both with train-time compilation (Section~\ref{training}) and ahead-of-time compilation for inference (Section~\ref{inference}).  \\

A key advantage of our TorchInductor approach is that it can accommodate dynamic tensor shapes, while many other neural network compilers enforce static tensor shapes.
In MLIP-based calculations---molecular dynamics in particular---both the number of atoms and the number of pairs of neighboring atoms input to the model can vary significantly over the course of a simulation.
TorchInductor generates a single set of optimized kernels that are applicable to the entire range of relevant dynamic shapes.

\subsection{Compiled and Distributed Training}
\label{training}

We apply TorchInductor to accelerate model training by using the procedure described in Section~\ref{inductor} to convert the entire model into a computational graph containing only primitive PyTorch operations.
We then apply \texttt{torch.compile} to this graph to generate optimized kernels for both the forward pass, which computes all of the model's predictions, and the train-time backward pass, which computes gradients of the loss function with respect to model parameters.  \\

To enable training on large datasets \citep{deng2023chgnet, schmidt2024improving, barroso2024open, kaplan2025foundational}, we have added distributed multi-GPU training to the NequIP framework.
We adopt a Distributed Data Parallel (DDP) paradigm \citep{li2020pytorch} via PyTorch Lightning \citep{Falcon_PyTorch_Lightning_2019}.
Instead of using PyTorch’s standard DDP implementation, which employs a gradient bucketing strategy that overlaps computation with communication, we implemented a custom DDP approach that performs communications only after the full backwards pass has completed.
This design choice allows TorchInductor to compile the entire backward pass, rather than splitting it into multiple subgraphs with communication operations in between, which can reduce overhead and introduce additional opportunities for fusion.
Practically, omitting gradient bucketing has not imposed meaningful limitations on our MLIP training, but a gradient bucketing strategy for larger models could be reintroduced in the future as needed.

\subsection{Ahead-of-Time Compilation for Inference}
\label{inference}

It is often important to integrate MLIP models with high-performance simulation codes that are written in low-level languages, such as the leading molecular dynamics code LAMMPS \citep{thompson2022CPC}, which is written in C++.
The NequIP framework previously interfaced with LAMMPS by exporting models to the TorchScript format and providing dedicated plugins \texttt{pair\_nequip} and \texttt{pair\_allegro} to load and call TorchScript-compiled models directly inside LAMMPS.
This approach has been adopted by other deep learning interatomic potential frameworks such as MACE \citep{batatia2022mace}, SevenNet \citep{park2024scalable}, and BAMBOO \citep{gong2024bamboo}.
The \texttt{pair\_allegro} plugin also uses the Kokkos performance portability library \citep{9485033} to reduce overhead in the interface between Allegro and LAMMPS by eliminating CPU–GPU data transfers wherever possible \citep{kozinsky2023scaling}. This feature also allows LAMMPS to use GPU-aware MPI communication. \\

Here, we introduce the first end-to-end use of Ahead-Of-Time Inductor (AOTI) compilation for MLIPs as a replacement for TorchScript export.
AOTI is a variant of TorchInductor specialized to export compiled PyTorch models as self-contained native code for use in non-Python environments. 
AOTI compilation allows us to realize the performance benefits of MLIP compilation in high-performance codes like LAMMPS without resorting to complicated, high-overhead solutions such as embedded Python interpreters \citep{devito2021using, wang2024openmm}.
Models compiled with AOTI can also still be used in Python-based codes like the Atomic Simulation Environment (ASE) \citep{ase-paper}.

\subsection{Optimized Tensor Product Kernel}
\label{kernel}

Custom kernels that implement fusion, scheduling, and memory-access patterns that are more advanced than those of current compilers have demonstrated significant success in accelerating deep learning models \citep{dao2022flashattention, hsu2024liger}.
Analogous efforts for equivariant neural networks have focused on optimizing various tensor product operations, which are a central computational bottleneck \citep{kovacs2023mace, batatia2023foundation, bharadwaj2025efficient, firoz2025optimizing}.
While Allegro’s tensor product operation already has a highly optimized pure PyTorch implementation \citep{kozinsky2023scaling}, it remains the most computationally expensive part of the Allegro architecture.
Profiling revealed that TorchInductor currently cannot fuse the entire pure PyTorch implementation of the Allegro tensor product into a single GPU kernel. Instead, it generates two separate kernels that require costly intermediate memory reads and writes. \\

To overcome the fusion limitations of TorchInductor, we implemented  fused custom kernels for the Allegro tensor product using Triton, a cross-platform language for writing high-performance compute kernels \citep{tillet2019triton}.
Because Triton is the native backend for TorchInductor, the kernel can be seamlessly integrated with our compilation infrastructure.
We implemented kernels for the tensor product and its first derivative to accelerate the inference-time prediction of energy and its first derivatives, like forces.
Importantly, custom Triton kernels integrate seamlessly with AOTI, ensuring that our custom kernel can be exported for use in LAMMPS and other inference software.
The additional kernels required for training are left for future work and would also straightforwardly integrate with train-time TorchInductor compilation.  \\

Our custom kernel takes advantage of the mathematical structure of the tensor product by using a compressed sparse format \citep{won2023unified} to represent the combined tensor of Wigner 3-$j$ contraction coefficients described in previous work \citep{kozinsky2023scaling}.
It also avoids materializing intermediate outer products between the input tensors, which results in significant memory savings.

\section{Case Study: Training and Inference of SPICE 2 Models}
\label{results}

\begin{figure*}[ht]
\centering
\includegraphics[width=0.95\linewidth]{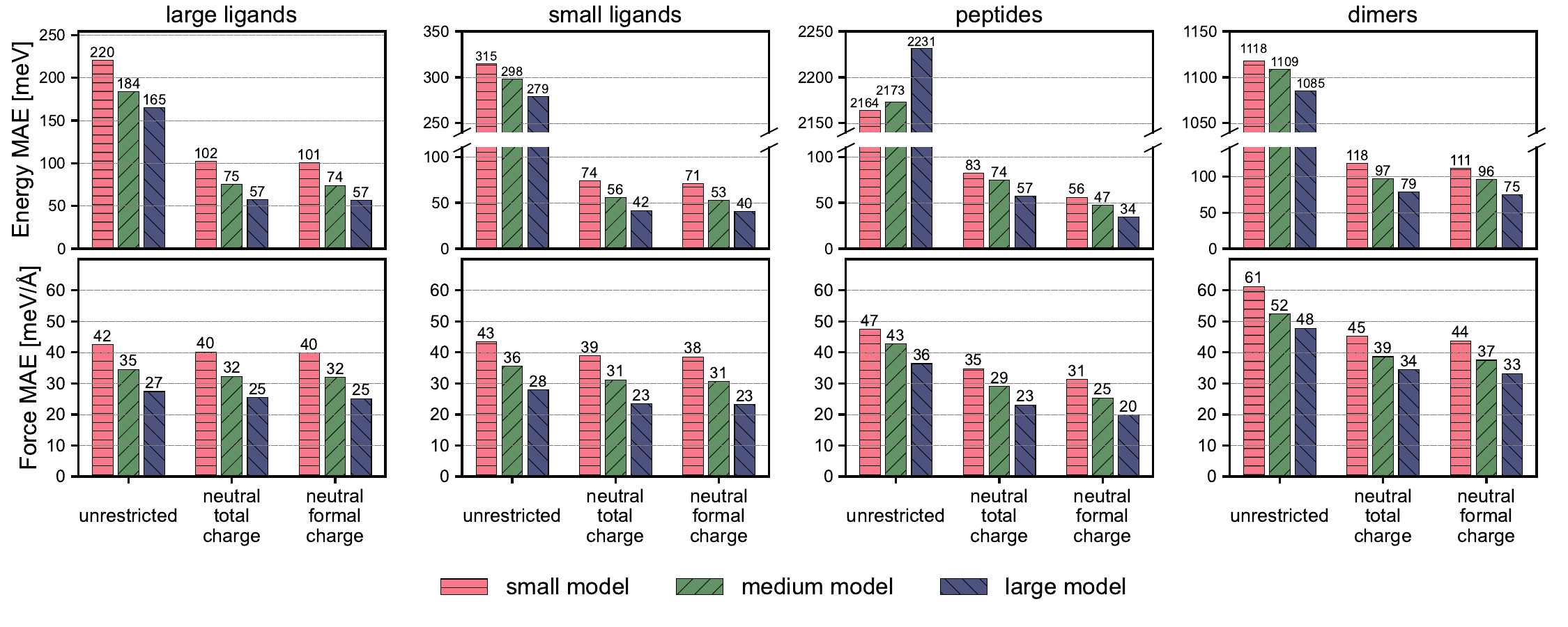}
\caption{\textbf{Allegro model results for the SPICE 2 test set} \citep{eastman2024nutmeg}. Energy and force mean absolute error (MAE) of the three Allegro models on three different subsets of the SPICE 2 test set: (1) the original unrestricted test set, (2) a subset containing systems with neutral total charge, and (3) a subset of systems that contain only atomic species with neutral per-atom formal charge. The error metrics are grouped by system type: large ligands, small ligands, peptides, and dimers.}
\label{fig:SPICE2-test}
\end{figure*}

To demonstrate the redesigned NequIP framework, we trained Allegro models on the SPICE 2 dataset \cite{eastman2023spice,eastman2024nutmeg} and benchmarked their performance in molecular dynamics.
This dataset contains more than two million atomic configurations of organic systems, including drug-like molecules, amino acids, ions, water clusters, and their interactions, with energies and forces calculated using density functional theory (DFT) with the $\omega$B97M-D3(BJ) functional \cite{najibi2018nonlocal,mardirossian2016omegab97m}.
Several general-purpose biomolecular MLIPs have been trained on the SPICE 2 data \citep{kovacs2023mace, hattori2024revisiting, eastman2024nutmeg}, leading to studies of the crystallization \citep{hattori2024revisiting} and hydration free energies \citep{moore2024computing} of small organic molecules with higher accuracy than classical force fields.
Accurate and efficient MLIPs for biomolecular systems are expected to have significant impact on drug discovery workflows \citep{barnett2024neural}. \\

The SPICE 2 dataset contains systems with a range of total charge states, which means that a given atomic configuration can have multiple possible DFT energy and force labels depending on the total charge used to compute them.
To avoid this degeneracy, the Allegro models in this work were trained on a subset of the SPICE 2 data limited to systems with neutral total charge.
This training strategy is less restrictive than that of \citet{kovacs2023mace}, who trained only on systems with neutral per-atom formal charge, but more restrictive than the approach of \citet{eastman2024nutmeg}, who trained on the full SPICE 2 dataset.  \\

Using this subset of SPICE 2, we prepared three Allegro models with different cost-accuracy trade-offs and will refer to them as ``small'', ``medium'', and ``large''.
The main differences between these models are the maximum rotation order ($\ell_\text{max}$),  the number of tensor features, and the number of Allegro layers, which are listed in Table~\ref{tab:SPICE-hparams}.
These hyperparameters control the cost and complexity of the Allegro tensor products that mix the models' equivariant features \citep{musaelian2023learning}.
All models employ a radial cutoff $r_\text{max}$ of 6.0~\si{\angstrom}, and
the Supplementary Information gives a full description of the hyperparameters and training procedure. 

\begin{table}[ht]
\def\arraystretch{1.25}
    \centering
    \caption{\textbf{Key hyperparameters} of the small, medium, and large Allegro models trained on the SPICE 2 dataset: the maximum rotation order ($\ell_\text{max}$), number of tensor features, and number of Allegro layers.}
    \label{tab:SPICE-hparams}
    \begin{tabular}{|c|c|c|c|}
    \hline
    Model & $\ell_\text{max}$ & Tensor Features & Allegro Layers \\
    \hline
    small & 2 & 64 & 2 \\
    \hline
    medium & 3 & 64 & 2 \\
    \hline
    large & 3 & 128 & 3 \\
    \hline
    \end{tabular}
\end{table}

\subsection{Model Accuracy}

To contextualize the differences in computational cost between the three models, we evaluate their accuracies using two benchmark datasets: the official SPICE 2 test set \citep{eastman2024nutmeg} and the TorsionNet 500 benchmark \citep{rai2022torsionnet} recomputed at the SPICE 2 level of theory  by \citet{eastman2024nutmeg}.  \\

Designed to assess the generalizability of MLIPs trained on SPICE 2, the SPICE 2 test set contains systems distinct from those in the training set, and it is divided into four categories: small ligands, large ligands, pentapeptides, and dimer pairs \citep{eastman2024nutmeg}.
To understand how our models (trained on neutral-total-charge data) perform on atomic configurations with different total- and per-atom- charge states, we evaluate the Allegro models on three subsets of the SPICE 2 test set: an unrestricted set that includes all systems in the test set regardless of charge state, a subset restricted to systems with neutral total charge (consistent with the data used to train the Allegro models), and a more restrictive subset of systems that contain only atomic species with neutral formal per-atom charge.
Figure~\ref{fig:SPICE2-test} shows a consistent trend across all system types and charge schemes where accuracy improves from the small to medium to large model.
While energy errors are large on the unfiltered test set, which includes molecular charge states that are entirely absent from the training data, the models perform well on both the neutral total charge subset and the neutral per-atom formal charge subset.
Force errors on the unfiltered test set are only slightly larger than on the two filtered subsets.  \\

The second benchmark dataset, TorsionNet 500, also considered by \citet{kovacs2023mace} and 
\citet{eastman2024nutmeg}, assesses an MLIP's ability to predict the relative energy differences between molecular conformers.
It contains 12,000 atomic configurations scanning through different values for the torsion angles along one bond in 500 drug-like molecules \citep{rai2022torsionnet}.
A key property of a torsion angle scan is the height of its energy barrier, which controls the likelihood of a conformational change involving that torsional angle.
Table~\ref{tab:SPICE-torsion} shows that the three Allegro models have barrier height errors comparable to those of \citet{kovacs2023mace} and 
\citet{eastman2024nutmeg}.  \\

Additional accuracy benchmarks for a held-out 5\% portion of the SPICE 2 training set, unused in our training, are presented in the Supplemental Information. \\

\begin{table}[t]
\def\arraystretch{1.25}
    \centering
    \caption{
    \textbf{Allegro model results for the TorsionNet 500 benchmark} \citep{rai2022torsionnet}. Mean absolute error (MAE) and root mean square error (RMSE) of the Allegro models' torsion barrier height predictions.
    }
    \label{tab:SPICE-torsion}
    \begin{tabular}{|c|c|c|c|}
    \hline
    Metric & small model & medium model & large model \\
    \hline
    Barrier MAE  ~~[meV] & 22.75 & 15.42 & 11.37 \\
    \hline
    Barrier RMSE [meV] & 32.36 & 21.77 & 15.38 \\
    \hline
    \end{tabular}
\end{table}

\begin{figure}[ht]
\centering
\includegraphics[width=0.95\linewidth]{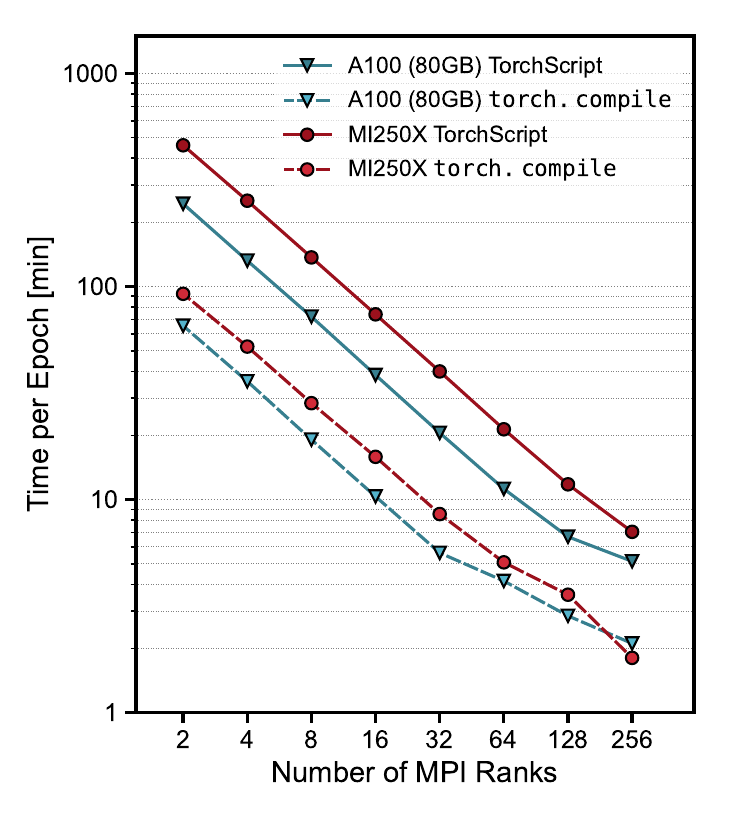}
\caption{\textbf{Scalability of distributed machine-learning interatomic potential training}. Average time per epoch of training across a range of numbers of MPI ranks, where the per-rank local batch size is eight atomic configurations. Times are measured for training with TorchScript or \texttt{torch.compile}. Plots are shown for both NVIDIA A100 (80GB) and AMD MI205X GPUs. Note that one MPI rank corresponds to one of the two available graphics compute dies on a single MI250X device.}
\label{fig:trainscale}
\end{figure}

\subsection{Train-time Acceleration}

\begin{figure*}[ht]
\centering
\includegraphics[width=0.95\linewidth]{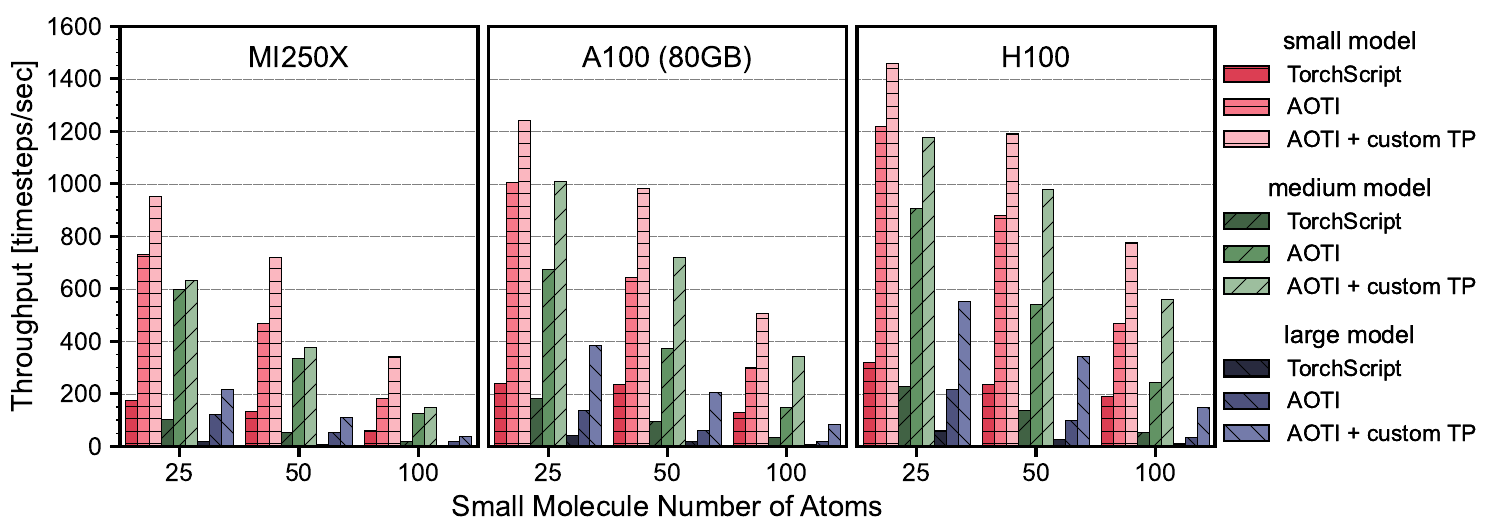}
\caption{\textbf{Single-rank inference acceleration on small molecule systems}. The inference speed in LAMMPS of the small, medium, and large Allegro models deployed using TorchScript, AOTI, or AOTI with the optimized tensor product kernel (AOTI + custom TP) on small molecule systems ranging from 25 to 100 atoms without periodic boundary conditions \citep{eastman2024nutmeg}, for AMD MI250X, NVIDIA A100 (80GB), and NVIDIA H100 GPUs. The inference speeds were averaged over three runs with different random seeds for the initial velocities generated by LAMMPS. One MPI rank was used. Note that one MPI rank corresponds to one of the two available graphics compute dies on an MI250X device.}
\label{fig:SPICE_SmallMol_Inference}
\end{figure*}

To demonstrate the efficiency and scalability of the redesigned NequIP infrastructure for  MLIP training, we compared the cost to train the medium-sized Allegro model on the SPICE 2 dataset using the previous TorchScript compiler and the new \texttt{torch.compile} in a distributed multi-GPU setting.
We performed the evaluation on both AMD MI250X GPUs using the Frontier supercomputer at the Oak Ridge Leadership Computing Facility (OLCF) and NVIDIA A100 GPUs using the Perlmutter supercomputer at the National Energy Research Scientific Computing Center (NERSC).
To reflect realistic training conditions, all metrics, callbacks, logging, and other sources of overhead typical of production training runs were included in the timings.  \\

Figure~\ref{fig:trainscale} presents the training performance as the number of MPI ranks increases, where each MPI rank runs a fixed per-rank local batch size of eight atomic configurations.
In this scaling regime, the effective total global batch size increases with the number of ranks, which allows the entire dataset to be processed (a single ``epoch'') in a smaller number of stochastic gradient descent steps, reducing the wall time per epoch.
Distributed training scales well up to 128 ranks.
At 256 ranks, it remains reasonably efficient and achieves parallel efficiencies of 40\% and 24\% on the AMD and NVIDIA systems, respectively, when running \texttt{torch.compile} and using the 2-rank performance as a baseline.
We observe that training with \texttt{torch.compile} achieves between 2.4–5.0 times speedups over TorchScript on MI250X and A100 (80GB) GPUs across the range of MPI ranks investigated.

\begin{figure*}[t]
\centering
\includegraphics[width=0.95\linewidth]{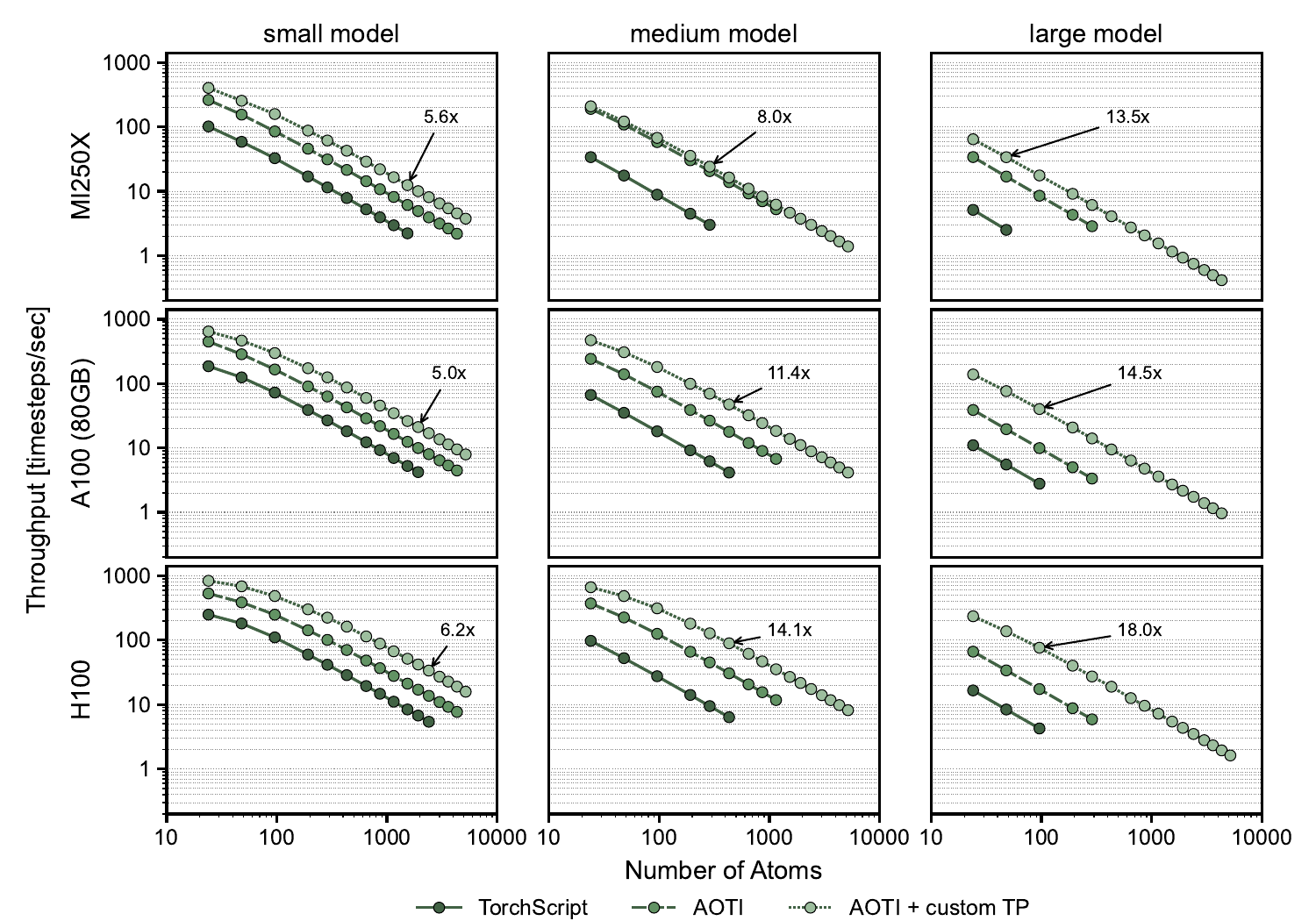}
\caption{\textbf{Single-rank inference acceleration for periodic water boxes}. Inference speed in LAMMPS for the small, medium, and large Allegro models deployed using TorchScript, AOTI, and AOTI with the optimized tensor product kernel (AOTI + custom TP) for liquid water boxes ranging from 24 to 5184 atoms with periodic boundary conditions. Annotations show the speedup of AOTI + custom TP compared to TorchScript for the largest system that both approaches can run. The speeds are measured on the AMD MI250X, NVIDIA A100 (80GB), and NVIDIA H100 GPUs. One MPI rank was used for each simulation (for the MI250X device, one MPI rank corresponds to one of the two available graphics compute dies).}
\label{fig:inference-water-1rank}
\end{figure*}

\begin{figure*}[ht]
\centering
\includegraphics[width=0.95\linewidth]{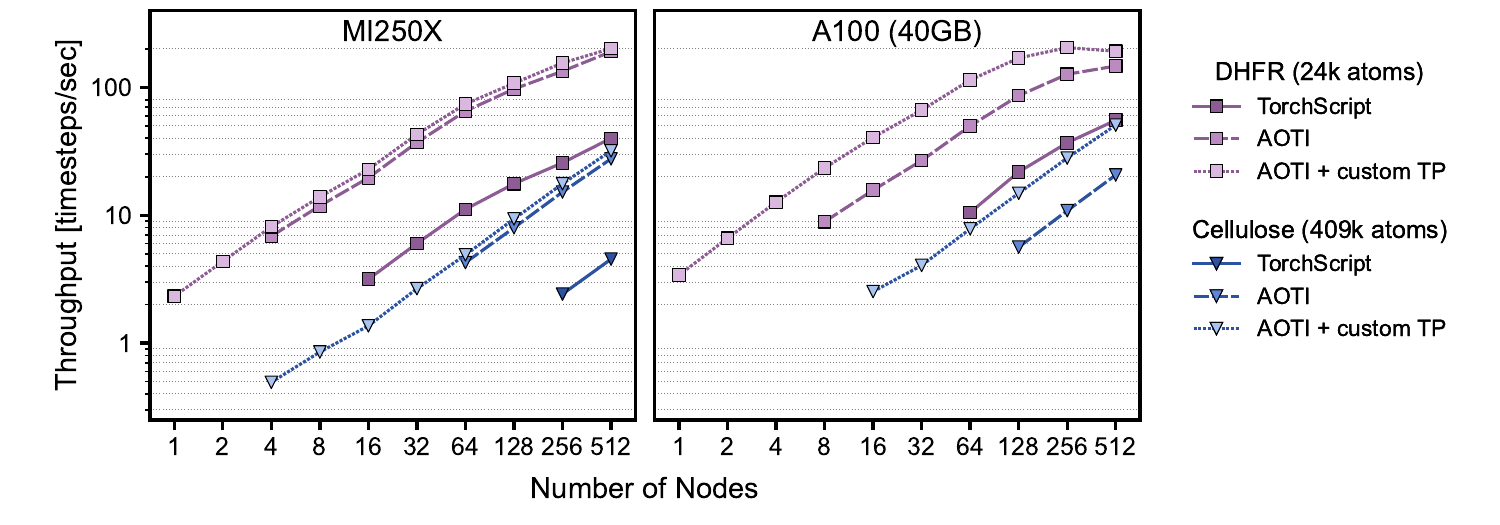}
\caption{\textbf{Strong scaling of the medium Allegro model on biomolecular systems}. The molecular dynamics throughput of the Allegro model deployed using TorchScript, AOTI, or AOTI with the optimized tensor product kernel (AOTI + custom TP) on the 23,558-atom dihydrofolate reductase (DHFR) and 408,609-atom cellulose systems from the Amber20 benchmark \citep{amberbench} is measured on a number of nodes ranging from 1 to 512 on Frontier (AMD MI250X) and Perlmutter (NVIDIA A100 (40GB)). Note that there are twice as many logical GPU devices and corresponding MPI ranks on an MI250X node (8) than on an A100 node (4). No TorchScript result is shown for cellulose on A100 GPUs because TorchScript required more GPU memory than was available on these nodes.}
\label{fig:inference-strong-scale}
\end{figure*}

\subsection{Inference Acceleration}
\label{inference-accel}

Next, we investigate the inference speed of the three Allegro models, highlighting how AOTI and the custom kernel provide dramatic improvements in throughput and memory efficiency.
To complete these experiments, we used the new \texttt{nequip-package} tool to portably archive the models, which were trained on Frontier, before compiling them on each inference platform using \texttt{nequip-compile}.
All inference benchmarks were performed in LAMMPS using our Kokkos-based integration for Allegro models.  \\

First, for a single AMD MI250X, NVIDIA A100 (80GB), and NVIDIA H100 GPU, we compare molecular dynamics performance for both small and large systems using: (1) small-molecule benchmarks (Figure~\ref{fig:SPICE_SmallMol_Inference}), introduced by \citet{eastman2024nutmeg}, consisting of 25-, 50- and 100-atom organic molecules in vacuum, and (2) periodic boxes of water (Figure~\ref{fig:inference-water-1rank}) at a density of 1 \si{g/cm^3} and temperature of 300 K with system sizes ranging from 24 to 5184 atoms.  \\

For all systems, AOTI exhibits a consistent performance advantage over TorchScript, and further speedups are achieved with the custom Triton tensor product.
The speedup factor tends to be greater for the larger Allegro models that have correspondingly more expensive tensor product operations for the custom kernel to accelerate.
Across all hardware and models, the small molecule examples (Figure~\ref{fig:SPICE_SmallMol_Inference}) see accelerations ranging from 4-18 times when comparing the TorchScript baseline to AOTI with the optimized TP; the same range of speedups is seen in the water simulations (Figure~\ref{fig:inference-water-1rank}).  \\

Figure~\ref{fig:inference-water-1rank} demonstrates that AOTI improves memory efficiency and that the the combination of AOTI and the optimized tensor product kernel dramatically increases the maximum number of atoms that can fit on one rank.
While the large model runs out of memory at water system sizes of around 100 atoms on all GPU types when using TorchScript, AOTI compilation extends this limit to 288 atoms.
The custom kernel then dramatically increases the maximum system size that the large model can run on one rank to 4320 atoms on an MI250X (specifically, one of the two available graphics compute dies), 4320 atoms on an A100 (80GB), and 5184 atoms on an H100.  \\

We also conducted multi-GPU strong scaling tests with the medium model on two all-atom water-solvated biomolecules: the 23,558-atom dihydrofolate reductase (DHFR) protein system and the 408,609-atom cellulose sugar polymer system, both from the Amber20 benchmark \citep{amberbench}.
Figure~\ref{fig:inference-strong-scale} shows excellent strong scaling up to 256 nodes on both benchmark systems when run on Frontier (AMD MI250X GPUs) and Perlmutter (NVIDIA A100 40GB GPUs).
Inference throughput on Perlmutter begins to saturate between 256 and 512 nodes for AOTI-compiled models (both with and without the custom kernel), while runs on Frontier continue to improve up to 512 nodes, where the AOTI compiled model using the custom kernel achieves over 200 timesteps per second.
We observe a maximum throughput of 205 timesteps per second on the 24k atom DHFR system using 256 nodes of Perlmutter with AOTI and the custom kernel.
The largest observed speedup over TorchScript is 10.9 times for DHFR on 64 nodes of Perlmutter.
For node counts where TorchScript inference fits in memory and can be run, we observe that the combination of AOTI and the kernel achieves an average speedup of 6.6 times on Frontier and 6.9 times on Perlmutter.  \\

Similar to Figure \ref{fig:inference-water-1rank}, the combination of AOTI compilation and the custom tensor product kernel dramatically improve memory efficiency.
For the larger cellulose system, the TorchScript-compiled model requires too much GPU memory for the 40GB A100s and fails to run even when using 512 nodes.
The AOTI and custom kernel approach, in contrast, runs successfully on as few as 16 nodes, which corresponds to approximately 6,384 atoms per GPU.  \\

\section{Conclusion}

We redesigned the NequIP framework to meet the increasing computational demands of MLIP training and inference. 
By integrating PyTorch 2.0 compiler technologies, distributed training, and optimized custom Triton kernels, the new infrastructure achieves substantial performance improvements for both training and inference.
We trained three Allegro models on the SPICE 2 dataset and demonstrated how the new infrastructure can support training MLIPs on large datasets.\\

The NequIP framework implements a number of techniques that could be applied to further accelerate the Allegro models presented here, and an in-depth study of their strengths and weaknesses is an important topic for future work.
In particular, the framework supports per-edge-type cutoff radii that depend on the atomic types of both the central and the neighbor atoms, which we used in \citep{kozinsky2023scaling}. 
Because the computational cost of Allegro scales linearly with the total number of edges, tuning per-edge-type cutoff radii can have a significant impact on speed.
The framework also supports the faster but less numerically accurate TensorFloat32 floating-point precision for intermediate computations, which we did not enable in this work.  \\

Finally, the updated NequIP framework is an extensible foundation for the development of MLIP architectures and training strategies.
The modular structure it enforces facilitates the development of extension packages like Allegro and ensures that they can easily leverage both the computational acceleration techniques described here and the broader set of tools, integrations, and infrastructure offered by the framework.

\section{Appendix}

\subsection{Mixed-Precision Architecture}

The Allegro model balances the performance benefits of lower-precision arithmetic with the need for numerical accuracy by employing a mix of double- and single-precision operations.
Specifically, the initial embedding of edge vectors between a central atom and its neighbors to a radial-chemical and spherical harmonics basis is performed largely in double-precision before casting the resulting embedded features to single-precision.
Subsequent operations in the Allegro layers, including the multilayer perceptrons (MLPs) acting on scalar features and the Clebsch–Gordan tensor products acting on tensor features, are carried out in single precision.
The learned representations are then passed through a readout MLP, also in single precision, and finally cast back to double precision when producing the energy prediction.
The backward pass used to compute forces and other derivative quantities follows this sequence in reverse, preserving the same precision transitions.

\subsection{Code}
This work used PyTorch version 2.6.0, \texttt{nequip} version 0.7.0 from \href{https://github.com/mir-group/nequip}{https://github.com/mir-group/nequip}, \texttt{allegro} version 0.4.0 from \href{https://github.com/mir-group/allegro}{https://github.com/mir-group/allegro}, and \texttt{pair\_nequip} version 0.7.0 from \href{https://github.com/mir-group/pair_nequip}{https://github.com/mir-group/pair\_nequip}.

\section{Acknowledgements}
We are grateful to Angela Yi and Richard Zhou for discussions regarding the PyTorch compilation workflow. We also thank Jaeyeon Won and Teodoro Collins for discussions regarding the custom kernel.  \\

This work was supported by the National Science Foundation through the Harvard University Materials Research Science and Engineering Center Grant No. DMR-2011754, the Department of Navy award N00014-20-1-2418 issued by the Office of Naval Research, and the Department of Energy Office of Basic Energy Sciences Award No. DE-SC0022199.
This work is also supported by Robert Bosch LLC and the National Science Foundation under Grant No. DMR-2119351.  \\

MK was supported by the National Science Foundation under Cooperative Agreement PHY-2019786 (The NSF AI Institute for Artificial Intelligence and Fundamental Interactions), and the NSF Graduate Research Fellowship program.
SRK thanks the Harvard University Center for the Environment (HUCE) for funding a fellowship. 
MW was supported by National Science Foundation, Office of Advanced Cyberinfrastructure (OAC), under Award No. 2118201.
AJ was supported by the Laboratory Directed Research and Development program at Sandia National Laboratories, a multimission laboratory managed and operated by National Technology and Engineering Solutions of Sandia, LLC, a wholly owned subsidiary of Honeywell International, Inc., for the U.S. Department of Energy’s National Nuclear Security Administration under contract DE-NA-0003525. This paper describes objective technical results and analysis. Any subjective views or opinions that might be expressed in the paper do not necessarily represent the views of the U.S. Department of Energy or the United States Government.  \\

An award of computer time was provided by the INCITE program. This research used resources of the Oak Ridge Leadership Computing Facility, which is a DOE Office of Science User Facility supported under Contract DE-AC05-00OR22725.

This research also used resources of the National Energy Research Scientific Computing Center (NERSC), a DOE Office of Science User Facility supported by the Office of Science of the U.S. Department of Energy under Contract No. DE-AC02-05CH11231 using NERSC awards BESERCAP0026720 and BESERCAP0032494. 

Additional computations were run on the FASRC Cannon cluster supported by the FAS Division of Science Research Computing Group at Harvard University.

%\nocite{*}
\bibliography{references}

\end{document}